\documentclass[12pt]{article}
\usepackage{amsmath}
\usepackage{graphicx}
\usepackage{float}
\newcommand{\be}{\begin{equation}}
\newcommand{\ee}{\end{equation}}
\newcommand{\ba}{\begin{eqnarray}}
\newcommand{\ea}{\end{eqnarray}}
\newcommand{\n}[1]{\label{#1}}

\begin{document}
\title{{\bf Approximations for the\\ Lowest Energy Eigenstates in a\\ Double Square Well Potential}\!\!
\thanks{Alberta-Thy-3-17, arXiv:yymm.nnnnn [hep-th]}}

\author{
Don N. Page\!\!
\thanks{Internet address:
profdonpage@gmail.com}
\\
Theoretical Physics Institute\\
Department of Physics\\
4-183 CCIS\\
University of Alberta\\
Edmonton, Alberta T6G 2E1\\
Canada
}

\date{2017 November 14}

\maketitle
\large
\begin{abstract}
\baselineskip 20 pt

Highly accurate closed-form approximations are given for the ground state and first excited state wavefunctions and energies for a nonrelativistic particle in a one-dimensional double square well potential with a square barrier in between (so that the potential is a sequence of five constant pieces that alternate in value from being above and below the ground state and first excited state energies), under the assumption that the barrier is sufficiently wide that the tunneling across it is very small.

\end{abstract}
\normalsize
\baselineskip 17 pt
\newpage

\section{Introduction}

One-dimensional double well potentials give relatively simple examples of quantum-mechanical tunneling and of the splitting of the energy eigenvalues for the ground state and the first excited state (symmetric and antisymmetric wavefunctions for a symmetric double well potential).  However, there appears to be no known elementary function form of a double well potential for which the ground state and first excited state wavefunctions can be calculated exactly in terms of elementary functions.  A quartic potential is perhaps the simplest form of a double well potential, but the energy eigenstates can only be calculated approximately or numerically.

Here I shall examine a double square well in somewhat more analytic detail than has been done previously \cite{JW,PL,JM,TMMP,DM}.  A double square well potential is piecewise constant (with five pieces and eight parameters, the values of the five potentials and the widths of the three finite regions that are the two wells and the barrier between them).  This potential has the advantage that for a given energy, the time-independent Schr\"{o}dinger equation implies that the wavefunction in each constant potential piece can be written explicitly in terms of hyperbolic and trigonometric sines and cosines with constant coefficients, though one gets transcendental equations to solve for the coefficients to match the wavefunction and its first derivative across the four boundaries between the five regions with the different constant potential values.  In the general case without symmetries, one gets four coupled transcendental equations to solve for four different parameters that are not determined by the general hyperbolic and trigonometric form of the wavefunction in each constant potential piece.  (The parameters to be solved can be the energy eigenvalue, the location of the maximum of the absolute value of the sinusoidal wavefunction in the left well, the location of the minimum of the absolute value of the hyperbolic wavefunction in the barrier between the two wells, and the location of the maximum of the absolute value of the sinusoidal wavefunction in the right well).

In this paper I assume that the barrier between the two wells is sufficiently high and/or broad that the hyperbolic wavefunction (for the ground state wavefunction, which without generality can have its phase chosen so that it is everywhere positive), or its derivative (for the first excited state wavefunction that has one sign change), changes by a large factor between its minimum inside the barrier and its value at each edge of the barrier.  This condition implies that the tunneling across the barrier has a large exponential suppression.  In this case one can get approximate starting-point solutions by assuming that the thick barrier is actually infinitely thick, so that the time-independent Schr\"{o}dinger equation decouples for the left and right sides, and for each side one gets one single transcendental equation (the root of a sum of two arcsines and a linear term in an initially unknown parameter whose square determines the energy) that can be readily solved on a pocket calculator, either by a series solution for certain small ranges of dimensionless combinations of the parameters of the potential, or else by Newton's method.

Then one can go back to the finitely thick barrier case as a first order perturbation in a calculable tunneling parameter to get both highly accurate approximate energy eigenvalues and eigenstates for the ground state and first excited state of the generic double square well potential.  These are thus given by an explicit algorithm (an explicit sequence of formulas, including iterations that in practice need be done only a very small number of times, say twice or thrice, in order to get results accurate to all the digits given by a pocket calculator).

One can use these explicit formulas first to examine a symmetric or nearly symmetric case in which the infinitely thick barrier limit gives equal energies for the decoupled ground state wavefunctions on the two sides.  In terms of the parameters of this infinitely thick barrier limit, one can readily calculate the energy splitting between the ground state and the first excited state when the barrier width is returned to its finite (but large) value, to first order in the tunneling parameter.  After this, motivated by the beautiful recent paper by Dauphinee and Marsiglio \cite{DM} of the large asymmetry in the wavefunction produced by a small asymmetry in the potential, I examine the case in which the potential is given a small perturbation (equal magnitudes but opposite signs for the potentials in the two wells, for simplicity leaving the potential values unchanged in the regions on the far left and far right and in the barrier in between the two wells, and also leaving unchanged the widths of all the potentials), so that in the decoupled infinitely thick barrier limit the ground state energies on the two sides become slightly different.  To lowest nontrivial order in this perturbation in the potential, one can then get explicit expressions for the ground state and first excited state energies and wavefunctions back in the finitely thick barrier case.  When the perturbation in the potential is of the order of the energy splitting between the ground state and the first excited state for the unperturbed potential, the asymmetry in the wavefunction becomes of the order of unity, though the exact value of the asymmetry of the probabilities for the particles to be on the two sides depends not only on the ratio of the potential perturbation to the energy splitting for the unperturbed potential but also on a coefficient of the order of unity that is calculated from the infinitely thick barrier limit and depends on the ratios of the potential differences inside and outside the well to the kinetic energy a particle would have inside the well if its walls had infinite potential.

\section{Ground State Wavefunction and Energy}

Let us consider a nonrelativistic particle of mass $m$ in a one-dimensional double square well potential
\ba
V &=& V_{-4},\ \ x<x_{-3}, \nonumber\\
V &=& V_{-2},\ \ x_{-3}<x<x_{-1}, \nonumber\\
V &=& V_{0},\ \ \,\ x_{-1}<x<x_1, \nonumber\\
V &=& V_{2},\ \ \,\ x_1<x<x_3, \nonumber\\
V &=& V_{4},\ \ \,\ x_3<x,
\n{pot}
\ea
where $V_{-4}>V_{-2}<V_0>V_2<V_4$ are constant potential values.  The two wells have potentials $V_{-2}$ (of width $w_{-2} = x_{-1}-x_{-3}$) and $V_2$ (of width $w_{2} = x_{3}-x_{1}$) and are surrounded on the left by $V_{-4}$ that extends to $x = -\infty$ and on the right by $V_4$ that extends to $x=+\infty$, and the two wells have a barrier of potential $V_0$ (of width $w_{0} = x_{1}-x_{-1}$) in between.  All of these parameters are assumed to be given and are assumed to be real.  It is also convenient to define positive potential differences between successive regions as
\ba
W_{-3} &\equiv& V_{-4}-V_{-2}, \nonumber\\
W_{-1} &\equiv& V_{0}-V_{-2}, \nonumber\\
W_{1} &\equiv& V_{0}-V_{2}, \nonumber\\
W_{3} &\equiv& V_{4}-V_{2}.
\ea

The main approximation in this paper is based on the assumption that the ground state wavefunction (with phase chosen to make it positive everywhere) will have a local minimum (at a location $x_0$ to be determined within the barrier) that is much smaller than the values of the wavefunction at both edges of the barrier, at $x_{-1}$ and $x_1$.  This requires that the ground state energy $E_0$ be less than the barrier potential $V_0$ in order that the wavefunction be concave upward inside the barrier so that it can have a minimum within the barrier.  Then for the wavefunction to avoid diverging as $x \rightarrow \pm\infty$, it must bend downward inside each well and thus have $E_0$ greater than the potentials $V_{-2}$ and $V_2$ in both wells.  Finally, in order for the wavefunction to be normalizable, it must tend to zero in the outer regions as $x \rightarrow \pm\infty$, so $E_0$ must be less than both $V_{-4}$ and $V_4$, so that the wavefunction does not oscillate in those regions of infinite width but instead is exponentially damped.  That is, we must have $V_{-4} > E_0 > V_{-2} < E_0 < V_0 > E_0 > V_2 < E_0 < V_4$.

I shall similarly assume that there is a first excited state wavefunction, $\tilde{\psi}(x)$, of energy $E_1 > E_0$, that is also chosen to be real but changes sign exactly once, at another location, $\tilde{x}_0$, that is between the edges of the barrier at $x_{-1}$ and $x_1$.  Therefore, $\tilde{\psi}(x)$ goes from being negative for $x < \tilde{x}_0$ to being positive for $x > \tilde{x}_0$.  For $\tilde{\psi}(x)$, the main assumption is that the slopes at the edges of the barrier at $x_{-1}$ and $x_1$, $\tilde{\psi}'(x_{-1})$ and $\tilde{\psi}'(x_1)$, are much greater than the slope at $\tilde{x}_0$, $\tilde{\psi}'(\tilde{x}_0)$.  For this wavefunction also to be finite and normalizable, one gets the analogous requirements that $V_{-4} > E_1 > V_{-2} < E_1 < V_0 > E_1 > V_2 < E_1 < V_4$.

First I shall focus on the ground state wavefunction $\psi(x)$.  By the time-independent Schr\"{o}dinger equation, it will have the following form in each of the five regions:
\ba
\psi(x) &=& \psi_{-4}(x) = A_{-4}e^{\kappa_{-4}(x-x_{-3})},\ \ x<x_{-3}, \nonumber\\
\psi(x) &=& \psi_{-2}(x) = A_{-2}\cos{[k_{-2}(x-x_{-2})]},\ \ x_{-3}<x<x_{-1}, \nonumber\\
\psi(x) &=& \psi_{0}(x)\ \ = A_0\cosh{[\kappa_0(x-x_0)]},\ \ x_{-1}<x<x_1, \nonumber\\
\psi(x) &=& \psi_{2}(x)\ \ = A_2\cos{[k_{2}(x-x_2)]},\ \ x_1<x<x_3, \nonumber\\
\psi(x) &=& \psi_{4}(x)\ \ = A_4e^{\kappa_{4}(x_3-x)},\ \ x_3<x,
\n{psi}
\ea
where
\ba
\kappa_{-4} &=& [2m(V_{-4}-E_0)]^{1/2}/\hbar, \nonumber\\
k_{-2} &=& [2m(E_0-V_{-2})]^{1/2}/\hbar, \nonumber\\
\kappa_{0} &=& [2m(V_{0}-E_0)]^{1/2}/\hbar, \nonumber\\
k_{2} &=& [2m(E_0-V_{2})]^{1/2}/\hbar, \nonumber\\
\kappa_{4} &=& [2m(V_{4}-E_0)]^{1/2}/\hbar,
\n{mom}
\ea
and where one also requires that the wavefunction $\psi(x)$ and its first derivative with respect to $x$, namely $\psi'(x)$, be continuous across the boundaries at $x_{-3}$, $x_{-1}$, $x_1$, and $x_3$, which gives the further conditions
\ba
A_{-4} &=& A_{-2}\cos{[k_{-2}(x_{-2}-x_{-3})]}, \nonumber\\
A_{-4}\kappa_{-4} &=& A_{-2}k_{-2}\sin{[k_{-2}(x_{-2}-x_{-3})]}, \nonumber\\
A_0\cosh{[\kappa_0(x_0-x_{-1})]} &=& A_{-2}\cos{[k_{-2}(x_{-1}-x_{-2})]}, \nonumber\\ 
A_0\kappa_{0}\sinh{[\kappa_0(x_0-x_{-1})]} &=& A_{-2}k_{-2}\sin{[k_{-2}(x_{-1}-x_{-2})]}, \nonumber\\
A_0\cosh{[\kappa_0(x_1-x_{0})]} &=& A_{2}\cos{[k_{2}(x_{2}-x_{1})]}, \nonumber\\ A_0\kappa_{0}\sinh{[\kappa_0(x_1-x_0)]} &=& A_0k_2\sin{[k_{2}(x_2-x_1)]}, \nonumber\\
A_{4} &=& A_{2}\cos{[k_{2}(x_{3}-x_{2})]}, \nonumber\\
A_{4}\kappa_{4} &=& A_{2}k_{2}\sin{[k_{2}(x_{3}-x_{2})]}.
\n{coef}
\ea

All of these parameters are can be chosen to be real and are (except for the given locations of the boundaries between the different values of the potential, at $x_{-3}$, $x_{-1}$, $x_1$, and $x_3$) to be determined from the given parameters by the time-independent Schr\"{o}dinger equation.  By the main assumption above, $\psi(x)$ has a local minimum within the barrier of potential $V_0$, at a position $x_{0}$ (to be determined) that is between the given locations $x_{-1}$ and $x_{1}$ of the edges of the barrier.  Then for $\psi(x)$ to bend back downward within the wells of potentials $V_{-2}$ and $V_2$, it must have a local maximum at a location $x_{-2}$ (to be determined) between the edges of the left well at $x_{-3}$ and $x_{-1}$ and another local maximum at a location $x_{2}$ (also to be determined) between the edges of the right well at $x_{1}$ and $x_{3}$.

The assumption that the local minimum of the wavefunction at the location $x_0$ within the barrier between the two wells is much less than the values of the wavefunction at the edges of the barrier at $x_{-1}$ and $x_1$ is the assumption that $\kappa_0(x_0-x_{-1})\gg 1$ and $\kappa_0(x_1-x_0)\gg 1$, so that $\cosh{[\kappa_0(x_0-x_{-1})]}\gg 1$ and $\cosh{[\kappa_0(x_1-x_0)]}\gg 1$.  

Subsidiary approximations that will be useful but not necessary in this paper will be that the ground state energy $E_0$ is much closer to $V_{-2}$ and $V_2$ (both of which $E_0$ must exceed in order that the ground state wavefunction have a local minimum within the barrier in between these two wells) than to $V_{-4}$, $V_0$, and $V_4$ (all three of which must be greater than $E_0$).

One can eliminate the coefficients $A_{-4}$, $A_{-2}$, $A_0$, $A_2$, and $A_4$ to get the following relations:
\ba
\tan{[k_{-2}(x_{-2}-x_{-3})]} &=& \kappa_{-4}/k_{-2}, \nonumber\\
\tan{[k_{-2}(x_{-1}-x_{-2})]} &=& (\kappa_0/k_{-2})\tanh{[\kappa_0(x_0-x_{-1})]}, \nonumber\\
\tan{[k_{2}(x_{2}-x_{1})]} &=& (\kappa_0/k_2)\tanh{[\kappa_0(x_1-x_0)]}, \nonumber\\
\tan{[k_{2}(x_{3}-x_{2})]} &=& \kappa_4/k_2.
\n{rel}
\ea

The arguments of the tangents are all less than $\pi/2$ radians (but are near that value when the $\kappa$'s are much greater than the $k$'s), so it is convenient to define
\ba
\varphi_{-3} &\equiv& \pi/2 - k_{-2}(x_{-2}-x_{-3}), \nonumber\\
\varphi_{-1} &\equiv& \pi/2 - k_{-2}(x_{-1}-x_{-2}), \nonumber\\
\varphi_{1} &\equiv& \pi/2 - k_{2}(x_{2}-x_{1}), \nonumber\\
\varphi_{3} &\equiv& \pi/2 - k_{2}(x_{3}-x_{2}).
\n{phi}
\ea

It is also convenient to define
\ba
K_{-2} &\equiv& \frac{\pi^2\hbar^2}{2m w_{-2}^2}, \nonumber\\
K_0 &\equiv& \frac{\pi^2\hbar^2}{2m w_0^2}, \nonumber\\
K_{2} &\equiv& \frac{\pi^2\hbar^2}{2m w_{2}^2},
\n{K}
\ea
which would be the ground state kinetic energies of a particle of mass $m$ confined respectively to regions of the widths $w_{-2}$, $w_0$, and $w_2$ of the three regions that have the potentials $V_{-2}$ (the left well), $V_0$ (the barrier between the wells), and $V_2$ (the right well), if in each case the potentials outside each of these regions were infinite rather than the actual values they have.  From these parameters determined directly from the particle mass $m$ and the original parameters of the double square well potential, one can get the following dimensionless parameters:
\ba
\alpha_{-3} &\equiv& \left(\frac{K_{-2}}{V_{-4}-V_{-2}}\right)^{1/2}
\equiv \left(\frac{K_{-2}}{W_{-3}}\right)^{1/2}, \nonumber\\
\alpha_{-1} &\equiv& \left(\frac{K_{-2}}{V_{0}-V_{-2}}\right)^{1/2}
\equiv \left(\frac{K_{-2}}{W_{-1}}\right)^{1/2}, \nonumber\\ 
\beta_{-1} &\equiv& \left(\frac{K_0}{V_{0}-V_{-2}}\right)^{1/2}
\equiv \left(\frac{K_0}{W_{-1}}\right)^{1/2}
\equiv \frac{w_{-2}}{w_0}\alpha_{-1}, \nonumber\\
\alpha_{1} &\equiv& \left(\frac{K_{2}}{V_{0}-V_{2}}\right)^{1/2}
\equiv \left(\frac{K_{2}}{W_{1}}\right)^{1/2}, \nonumber\\ 
\beta_{1} &\equiv& \left(\frac{K_0}{V_{0}-V_{2}}\right)^{1/2}
\equiv \left(\frac{K_0}{W_1}\right)^{1/2}
\equiv \frac{w_{2}}{w_0}\alpha_{1}, \nonumber\\
\alpha_{3} &\equiv& \left(\frac{K_{2}}{V_{4}-V_{2}}\right)^{1/2}
\equiv \left(\frac{K_{2}}{W_{3}}\right)^{1/2}.
\n{alpha}
\ea
These also are directly determined by the mass and the original parameters of the potential (both the potential values of the different regions and their widths).

Now define
\ba
y_{-2} &\equiv& \left(\frac{E_0-V_{-2}}{K_{-2}}\right)^{1/2}, \nonumber\\
y_{2} &\equiv& \left(\frac{E_0-V_{2}}{K_{2}}\right)^{1/2},
\n{y}
\ea
whose determination is equivalent to the determination of the ground state energy
\be 
E_0 = V_{-2} + K_{-2}y_{-2}^2 = V_{2} + K_{2}y_{2}^2.
\n{E} 
\ee
In terms of these values $y_{-2}$ and $y_2$, also define
\ba
s_{-3} &\equiv& \alpha_{-3} y_{-2}, \nonumber\\
s_{-1} &\equiv& \alpha_{-1} y_{-2}, \nonumber\\
s_{1} &\equiv& \alpha_{1} y_{2}, \nonumber\\
s_{3} &\equiv& \alpha_{3} y_{2}.
\n{s}
\ea
It is furthermore useful to define
\ba
r_{-1} &\equiv& \kappa_0(x_0-x_{-1}), \nonumber\\
r_{1} &\equiv& \kappa_0(x_1-x_{0}), \nonumber\\
r_0 &\equiv& r_{-1} + r_1 \equiv \kappa_0(x_1-x_{-1}) \equiv \kappa_0 w_0.
\n{r}
\ea

Then in terms of these quantities, Eqs.\ (\ref{rel}) that come from the matching of $\psi(x)$ and $\psi'(x)$ across the boundaries between the different values of the potential become
\ba
\tan{\varphi_{-3}} &=& k_{-2}/\kappa_{-4}, \nonumber\\
\tan{\varphi_{-1}} &=& (k_{-2}/\kappa_{0})\coth{r_{-1}}, \nonumber\\
\tan{\varphi_{1}} &=& (k_{2}/\kappa_{0})\coth{r_{1}}, \nonumber\\
\tan{\varphi_{3}} &=& k_{2}/\kappa_{4}.
\n{tan}
\ea
These equations combined with Eqs.\ (\ref{mom}) and the other definitions above then give the following set of equations:
\ba
\sin^{-1}{(s_{-3})} + \sin^{-1}{\left(\frac{s_{-1}}{\sqrt{1-(1-s_{-1}^2)/\cosh^2{(r_{-1})}}}\right)} = \pi - \pi y_{-2}, \nonumber\\
\sin^{-1}{(s_{3})} + \sin^{-1}{\left(\frac{s_{1}}{\sqrt{1-(1-s_{1}^2)/\cosh^2{(r_{1})}}}\right)} = \pi - \pi y_{2}, \nonumber\\
r_{-1} + r_1 = \frac{\pi}{\beta_{-1}}\sqrt{1-s_{-1}^2} = \frac{\pi}{\beta_{1}}\sqrt{1-s_{1}^2}.
\n{set}
\ea

The 8 Eqs.\ (\ref{s}) and (\ref{set}) then determine the 8 unknowns, $y_{-2}$, $y_2$, $s_{-3}$, $s_{-1}$, $s_1$, $s_3$, $r_{-1}$, and $r_1$, in terms of the known dimensionless parameters $\alpha_{-3}$, $\alpha_{-1}$, $\beta_{-1}$, $\alpha_1$, $\beta_1$, and $\alpha_3$ that are given by Eqs.\ (\ref{K}) and (\ref{alpha}) directly in terms of the particle mass and the form of the potential.  One can easily use Eqs.\ (\ref{s}) to eliminate $s_{-3}$, $s_{-1}$, $s_1$, and $s_3$ from Eqs.\ (\ref{set}) so that they have only the 4 unknowns $y_{-2}$, $y_2$, $r_{-1}$, and $r_1$, but perhaps Eqs.\ (\ref{set}) are a bit more clear in the form I have written them here.

Now to get highly accurate approximate solutions to Eqs.\ (\ref{set}), use the assumption that $r_{-1} \gg 1$ and $r_1 \gg 1$, so that $1/\cosh^2{(r_{-1})} \approx 4e^{-2r_{-1}} \ll 1$ and $1/\cosh^2{(r_{1})} \approx 4e^{-2r_{1}} \ll 1$.  Then I shall solve Eqs.\ (\ref{set}) to first order in $e^{-2r_{-1}}$ and $e^{-2r_{1}}$.

First it is convenient to solve the first two equations of Eqs.\ (\ref{set}) to zeroth order in $e^{-2r_{-1}}$ and $e^{-2r_{1}}$.  Let me use $Y_{-2}$ and $Y_2$ for the values of $y_{-2}$ and $y_2$ that solve these equations when one sets $1/\cosh^2{(r_{-1})} = 1/\cosh^2{(r_{1})} = 0$:
\ba
\sin^{-1}{(\alpha_{-1} Y_{-2})}+\sin^{-1}{(\alpha_{-3} Y_{-2})} &=& \pi - \pi Y_{-2}, \nonumber\\
\sin^{-1}{(\alpha_{1} Y_{2})} + \sin^{-1}{(\alpha_{3} Y_{2})} &=& \pi - \pi Y_{2}.
\n{Y}
\ea

These are the equations that would apply exactly if the width $w_0$ of the barrier between the two wells, with potential $V_0$, were taken to infinity, so that there would be no tunneling between the two wells with potentials $V_{-2}$ and $V_2$, and so that their ground state energies and wavefunctions would be determined independently of each other.  This is what I shall call the infinitely thick barrier limit.

In these equations, the arcsines need to be between 0 and $\pi/2$.  Letting $Y$ denote either $Y_{-2}$ or $Y_2$, $\alpha_+$ the larger of the two corresponding $\alpha$'s, and $\alpha_-$ the smaller of the two corresponding $\alpha$'s, then the relevant solution has $Y \leq 1$ and $Y \leq 1/\alpha_+$.  If $\alpha_+ \leq 2$, then there is a solution for $Y$ for all $\alpha_- \leq \alpha_+$.  However, if $\alpha_+ > 2$, one needs $\alpha_- \geq \alpha_+\cos{(\pi/\alpha_+)}$ for there to be a solution, a bound state in the well in the infinitely thick barrier limit.  Note that $\alpha_- = \alpha_+$ always gives a solution for any $\alpha_+$, because then the potential on both sides of the well is the same, and any potential well in an otherwise constant potential always gives a bound state in one dimension.

If the $\alpha$'s are sufficiently small, one can get enough terms of a series solution to get the values of $Y_{-2}$ and $Y_2$ to high accuracy.  For example, using the Taylor expansion for the arcsine, one can write
\ba
\pi - \pi Y_{2} = \alpha_{1} Y_{2} + \frac{1}{6}(\alpha_{1} Y_{2})^3 + \frac{3}{40}(\alpha_{1} Y_{2})^5 + \frac{5}{112}(\alpha_{1} Y_{2})^7 + 
\frac{35}{1152}(\alpha_{1} Y_{2})^9 + \cdots \nonumber\\
+ \alpha_{3} Y_{2} + \frac{1}{6}(\alpha_{3} Y_{2})^3 + \frac{3}{40}(\alpha_{3} Y_{2})^5 + \frac{5}{112}(\alpha_{3} Y_{2})^7 + 
\frac{35}{1152}(\alpha_{3} Y_{2})^9 + \cdots,
\ea
or
\ba
Y_2 = \frac{\pi}{\pi+\alpha_{1}+\alpha_{3}}
[1-\frac{1}{6\pi}(\alpha_{1}^3+\alpha_{3}^3)Y_2^3
-\frac{3}{40\pi}(\alpha_{1}^5+\alpha_{3}^5)Y_2^5 \nonumber\\
-\frac{5}{112\pi}(\alpha_{1}^7+\alpha_{3}^7)Y_2^7
-\frac{35}{1152\pi}(\alpha_{1}^9+\alpha_{3}^9)Y_2^9-\cdots].
\ea
Then if one defines
\ba
\gamma_{-3} &\equiv& \frac{\pi\alpha_{-3}}{\pi+\alpha_{-3}+\alpha_{-1}}, \nonumber\\
\gamma_{-1} &\equiv& \frac{\pi\alpha_{-1}}{\pi+\alpha_{-3}+\alpha_{-1}}, \nonumber\\
\gamma_{1} &\equiv& \frac{\pi\alpha_{1}}{\pi+\alpha_{1}+\alpha_{3}}, \nonumber\\
\gamma_{3} &\equiv& \frac{\pi\alpha_{3}}{\pi+\alpha_{1}+\alpha_{3}},
\ea
one gets
\ba
Y_{-2} = \frac{\pi}{\pi+\alpha_{-1}+\alpha_{-3}}
[1-\frac{1}{6\pi}(\gamma_{-1}^3+\gamma_{-3}^3)
-\frac{3}{40\pi}(\gamma_{-1}^5+\gamma_{-3}^5)
+\frac{1}{12\pi^2}(\gamma_{-1}^3+\gamma_{-3}^3)^2 \nonumber\\
-\frac{5}{112\pi}(\gamma_{-1}^7+\gamma_{-3}^7)
+\frac{1}{10\pi^2}(\gamma_{-1}^3+\gamma_{-3}^3)(\gamma_{-1}^5+\gamma_{-3}^5)
\nonumber\\
-\frac{35}{1152\pi}(\gamma_{-1}^9+\gamma_{-3}^9)
-\frac{1}{18\pi^3}(\gamma_{-1}^3+\gamma_{-3}^3)^3 \nonumber\\
+\frac{25}{336\pi^2}(\gamma_{-1}^3+\gamma_{-3}^3)(\gamma_{-1}^7+\gamma_{-3}^7)
+\frac{9}{320\pi^2}(\gamma_{-1}^5+\gamma_{-3}^5)^2 +\cdots], \nonumber\\
Y_2 = \frac{\pi}{\pi+\alpha_{1}+\alpha_{3}}
[1-\frac{1}{6\pi}(\gamma_{1}^3+\gamma_{3}^3)
-\frac{3}{40\pi}(\gamma_{1}^5+\gamma_{3}^5)
+\frac{1}{12\pi^2}(\gamma_{1}^3+\gamma_{3}^3)^2 \nonumber\\
-\frac{5}{112\pi}(\gamma_{1}^7+\gamma_{3}^7)
+\frac{1}{10\pi^2}(\gamma_{1}^3+\gamma_{3}^3)(\gamma_{1}^5+\gamma_{3}^5)
\nonumber\\
-\frac{35}{1152\pi}(\gamma_{1}^9+\gamma_{3}^9)
-\frac{1}{18\pi^3}(\gamma_{1}^3+\gamma_{3}^3)^3 \nonumber\\
+\frac{25}{336\pi^2}(\gamma_{1}^3+\gamma_{3}^3)(\gamma_{1}^7+\gamma_{3}^7)
+\frac{9}{320\pi^2}(\gamma_{1}^5+\gamma_{3}^5)^2 +\cdots].
\n{Yseries}
\ea
Even if one extended this truncated series to an infinite series, it would not converge unless all the $\gamma$'s were less than unity, so the series above should be used only if all the $\gamma$'s are sufficiently less than unity that this truncated series gives a good approximation.

An alternative way to solve Eqs.\ (\ref{Y}) that does not require that the $\alpha$'s be so small is to use Newton's method with initial approximate solutions, such as $Y_{-2(0)} = 1$ or $Y_{2(0)} = 1$ if the corresponding $\alpha_+ < 1$, or $Y_{-2(0)}$ and/or $Y_{2(0)}$ just a bit smaller than the corresponding $1/\alpha_+$ if this is less than one.  If both $\gamma_{-1}^3+\gamma_{-3}^3$ and $\gamma_{1}^3+\gamma_{3}^3$ are sufficiently smaller than $6\pi$, perhaps a better first estimate would be
\ba
Y_{-2(1)} &=& \frac{\pi}{\pi+\alpha_{-1}+\alpha_{-3}}
[1-\frac{1}{6\pi}(\gamma_{-1}^3+\gamma_{-3}^3)], \nonumber\\
Y_{2(1)} &=& \frac{\pi}{\pi+\alpha_{1}+\alpha_{3}}
[1-\frac{1}{6\pi}(\gamma_{1}^3+\gamma_{3}^3)].
\n{Yest1}
\ea
Then with whatever initial $Y_{-2(0)}$ and $Y_{2(0)}$ are appropriate, iterate
\ba
Y_{-2(n+1)}\!\! &=&\!\! \frac{\pi\! +\! \frac{\alpha_{-1}Y_{-2(n)}}{\sqrt{1-\alpha_{-1}^2Y_{-2(n)}^2}}
\!+\!\frac{\alpha_{-3}Y_{-2(n)}}{\sqrt{1-\alpha_{-3}^2Y_{-2(n)}^2}}
\!-\!\sin^{-1}{(\alpha_{-1}Y_{-2(n)})}\!-\!\sin^{-1}{(\alpha_{-3}Y_{-2(n)})}}
{\pi + \frac{\alpha_{-1}}{\sqrt{1-\alpha_{-1}^2Y_{-2(n)}^2}}
+\frac{\alpha_{-3}}{\sqrt{1-\alpha_{-3}^2Y_{-2(n)}^2}}}, \nonumber\\
Y_{2(n+1)}\!\! &=&\!\! \frac{\pi + \frac{\alpha_{1}Y_{2(n)}}{\sqrt{1-\alpha_{1}^2Y_{2(n)}^2}}
+\frac{\alpha_{3}Y_{2(n)}}{\sqrt{1-\alpha_{3}^2Y_{2(n)}^2}}
-\sin^{-1}{(\alpha_{1}Y_{2(n)})}-\sin^{-1}{(\alpha_{3}Y_{2(n)})}}
{\pi + \frac{\alpha_{1}}{\sqrt{1-\alpha_{1}^2Y_{2(n)}^2}}
+\frac{\alpha_{3}}{\sqrt{1-\alpha_{3}^2Y_{2(n)}^2}}}.
\n{Yestn}
\ea

After solving for $Y_{-2}$ and $Y_2$, define, analogous to Eqs.\ (\ref{s}),
\ba
S_{-3} &\equiv& \alpha_{-3} Y_{-2}, \nonumber\\
S_{-1} &\equiv& \alpha_{-1} Y_{-2}, \nonumber\\
S_{1} &\equiv& \alpha_{1} Y_{2}, \nonumber\\
S_{3} &\equiv& \alpha_{3} Y_{2}.
\n{S}
\ea
It is also convenient further to define
\ba
\Phi_{-3} &\equiv& \sin^{-1}{(S_{-3})} \equiv \sin^{-1}{(\alpha_{-3} Y_{-2})}, \nonumber\\
\Phi_{-1} &\equiv& \sin^{-1}{(S_{-1})} \equiv \sin^{-1}{(\alpha_{-1} Y_{-2})}, \nonumber\\
\Phi_{1} &\equiv& \sin^{-1}{(S_{1})} \equiv \sin^{-1}{(\alpha_{1} Y_{2})}, \nonumber\\
\Phi_{3} &\equiv& \sin^{-1}{(S_{3})} \equiv \sin^{-1}{(\alpha_{3} Y_{2})}, 
\n{Phi}
\ea
\ba
C_{-3} &\equiv& \cos{\Phi_{-3}} \equiv \sqrt{1-S_{-3}^2}, \nonumber\\
C_{-1} &\equiv& \cos{\Phi_{-1}} \equiv \sqrt{1-S_{-1}^2}, \nonumber\\
C_{1} &\equiv& \cos{\Phi_{1}} \equiv \sqrt{1-S_{1}^2}, \nonumber\\
C_{3} &\equiv& \cos{\Phi_{3}} \equiv \sqrt{1-S_{3}^2},
\n{C}
\ea
\ba
T_{-3} &\equiv& \tan{\Phi_{-3}} \equiv \frac{S_{-3}}{\sqrt{1-S_{-3}^2}}, \nonumber\\
T_{-1} &\equiv& \tan{\Phi_{-1}} \equiv \frac{S_{-1}}{\sqrt{1-S_{-1}^2}}, \nonumber\\
T_{1} &\equiv& \tan{\Phi_{1}} \equiv \frac{S_{1}}{\sqrt{1-S_{1}^2}}, \nonumber\\
T_{3} &\equiv& \tan{\Phi_{3}} \equiv \frac{S_{3}}{\sqrt{1-S_{3}^2}}.
\n{T}
\ea

Then if one writes
\ba
y_{-2} &\equiv& Y_{-2}(1-\epsilon_{-2}), \nonumber\\
y_{2} &\equiv& Y_{2}(1-\epsilon_{2}),
\n{yY}
\ea
the first two of Eqs.\ (\ref{set}) imply that to first order in $e^{-2r_{-1}}$ and $e^{-2r_{1}}$,
\ba
\epsilon_{-2} &\approx& c_{-2}e^{-2r_{-1}}, \nonumber\\
\epsilon_{2} &\approx& c_{2}e^{-2r_{1}},
\n{eps}
\ea
where
\ba
c_{-2} &\equiv& \frac{2S_{-1}C_{-1}}{T_{-3}+T_{-1}+\pi Y_{-2}}
\equiv \frac{2}{\pi}S_{-1}C_{-1}U_{-2}, \nonumber\\
c_{2} &\equiv& \frac{2S_{1}C_{1}}{T_{1}+T_{3}+\pi Y_{2}}
\equiv \frac{2}{\pi}S_{1}C_{1}U_{2},
\n{c}
\ea
with
\ba
U_{-2} &\equiv& \frac{\pi}{T_{-3}+T_{-1}+\pi Y_{-2}}, \nonumber\\
U_{2} &\equiv& \frac{\pi}{T_{1}+T_{3}+\pi Y_{2}}.
\ea

If now one defines
\ba
a_{-1} &\equiv& \frac{\pi C_{-1}}{\beta_{-1}}, \nonumber\\
b_{-1} &\equiv& \frac{\pi S_{-1} T_{-1}}{\beta_{-1}} \equiv T_{-1}^2 a_{-1}, \nonumber\\
a_{1} &\equiv& \frac{\pi C_{1}}{\beta_{1}}, \nonumber\\
b_{1} &\equiv& \frac{\pi S_{1} T_{1}}{\beta_{1}} \equiv T_{1}^2 a_{1}, \nonumber\\
P &\equiv& b_{-1}b_1c_{-2}c_2
\equiv \frac{4}{\pi^2}a_{-1}a_1S_{-1}^2S_1^2T_{-1}T_1U_{-2}U_2
\equiv \frac{4}{\beta_{-1}\beta_1}S_{-1}^3S_1^3U_{-2}U_2,
\n{ab}
\ea
then the second two of Eqs.\ (\ref{set}) imply that to first order in $e^{-2r_{-1}}$ and $e^{-2r_{1}}$,
\ba
r_0 \equiv r_{-1} + r_1 &\approx& a_{-1}+b_{-1}\epsilon_{-2} \approx a_{1}+b_{1}\epsilon_{2} \nonumber\\
&\approx& a_{-1}+b_{-1}c_{-2}e^{-2r_{-1}} \approx a_{1}+b_{1}c_{2}e^{-2r_{1}}.
\n{reps}
\ea
Then
\ba
p \equiv P e^{-2r_{0}} \approx (r_0-a_{-1})(r_0-a_1),
\n{p}
\ea
so
\ba
r_0 \approx \frac{a_{1}+a_{-1}}{2} + \sqrt{\left(\frac{a_{1}-a_{-1}}{2}\right)^2 + p}\,.
\n{r_0}
\ea

This equation does not directly give $r_0$, since $p$ on the right hand side depends on $r_0$ as given by Eq.\ (\ref{p}), but one may iterate to get a highly accurate approximation for $r_0$.  For example, since we are assuming $r_0 \gg 1$ in Eq.\ (\ref{p}), we can get as a first approximation that $p=0$ and insert this into the right hand side of Eq.\ (\ref{r_0}) to get as a first approximation that $r_0$ is the maximum of $a_{-1}$ and $a_1$, and then using this first approximation for $r_0$ back in Eq.\ (\ref{p}) gives a second approximation for $p$, which one can insert back into Eq.\ (\ref{r_0}) to get a second approximation for $r_0$, and so on until these approximations converge sufficiently, as they generally do quite rapidly. 

After getting a sufficiently good approximation for $r_0$, one can then use Eq.\ (\ref{reps}) to get
\ba
\epsilon_{-2} &\approx& \frac{r_0-a_{-1}}{b_{-1}}
\approx \frac{1}{2b_{-1}}\left[(a_1-a_{-1})+\sqrt{(a_1-a_{-1})^2 + 4p}\right], \nonumber\\ 
\epsilon_{2} &\approx& \frac{r_0-a_{1}}{b_{1}}
\approx \frac{1}{2b_{1}}\left[-(a_{1}-a_{-1})+\sqrt{(a_{1}-a_{-1})^2 + 4p}\right].
\ea
Then one can plug these back into Eqs.\ (\ref{yY}) to get $y_{-2}$ and $y_2$ and then insert these into Eqs.\ (\ref{E}) to get two estimates for the ground state energy $E_0$, which should be very close to each other and to the exact ground state energy if $r_{-1}\gg 1$ and $r_1\gg 1$.

Now when the potential is not precisely symmetric about the middle of the barrier, we would like the asymmetry in the coefficients $A_{-2}$ and $A_2$ of the wavefunction in the two wells.  If one uses Eqs.\ (\ref{coef}), (\ref{phi}), and (\ref{r}), and the zeroth-order (in $e^{-2r_{-1}}$ and $e^{-2r_{1}}$) approximations
\ba
\sin{\varphi_{-3}} &\approx& s_{-3} \equiv \alpha_{-3} y_{-2} \equiv \alpha_{-3} Y_{-2}(1-\epsilon_{-2}) \approx \alpha_{-3} Y_{-2} \equiv S_{-3}, \nonumber\\
\sin{\varphi_{-1}} &\approx& s_{-1} \equiv \alpha_{-1} y_{-2} \equiv \alpha_{-1} Y_{-2}(1-\epsilon_{-2}) \approx \alpha_{-1} Y_{-2} \equiv S_{-1}, \nonumber\\
\sin{\varphi_{1}} &\approx& s_{1} \equiv \alpha_{1} y_{2} \equiv \alpha_{1} Y_{2}(1-\epsilon_{2}) \approx \alpha_{1} Y_{2} \equiv S_{1}, \nonumber\\
\sin{\varphi_{3}} &\approx& s_{3} \equiv \alpha_{3} y_{2} \equiv \alpha_{3} Y_{2}(1-\epsilon_{2}) \approx \alpha_{3} Y_{2} \equiv S_{3},
\ea
we get
\be
\frac{A_{-2}}{A_2} = \frac{\cosh{r_{-1}\sin{\varphi_{-1}}}}{\cosh{r_{1}\sin{\varphi_{1}}}} \approx \frac{S_1}{S_{-1}}e^{r_{-1}-r_1}.
\ee
Eqs.\ (\ref{eps}), (\ref{p}), and (\ref{r_0}) imply that
\be
e^{2(r_{-1}-r_1)} \approx \frac{c_{-2}\epsilon_2}{c_2\epsilon_{-2}} \approx
\left[\frac{b_{-1}c_{-2}}{b_1c_2}\right]
\left[\frac{-(a_{1}-a_{-1})+\sqrt{(a_{1}-a_{-1})^2 + 4p}}
{(a_{1}-a_{-1})+\sqrt{(a_{1}-a_{-1})^2 + 4p}}\right],
\ee
so
\be
\left(\frac{A_{-2}}{A_2}\right)^2 \approx 
\left[\frac{S_1^2 b_{-1}c_{-2}}{S_{-1}^2 b_1c_2}\right]
\left[\frac{-(a_{1}-a_{-1})+\sqrt{(a_{1}-a_{-1})^2 + 4p}}
{(a_{1}-a_{-1})+\sqrt{(a_{1}-a_{-1})^2 + 4p}}\right].
\n{coefrat}
\ee

When one goes back to the limit of an infinitely thick barrier (with potential $V_0$) between the left well (with potential $V_{-2}$) and the right well (with potential $V_2$), then with a real ground state wavefunction for each of the two wells (each of which exponentially decays with distance as it extends into the potential $V_{-4}$ at the extreme left side or $V_4$ at the extreme right side, and also into the potential $V_0$ of the barrier in between), one gets that the probabilities given by the wavefunctions $\psi_L$ and $\psi_R$ on the two sides are
\ba
P_L &=& \int |\psi_L|^2 dx \approx \frac{1}{2}\frac{A_{-2}^2 w_{-2}}{U_{-2} Y_{-2}}, \nonumber\\
P_R &=& \int |\psi_R|^2 dx \approx \frac{1}{2}\frac{A_{2}^2 w_{2}}{U_{2} Y_{2}}.
\n{P}
\ea

These equations are exact in the limit of an infinitely thick barrier between the two wells (so that neither $\psi_L$ nor $\psi_R$ has any leakage into the other well), but they are only approximate in the situation actually under consideration (a finitely thick barrier), in which there is a large but not infinite exponential decay of the wavefunction with distance from the wells inside the central barrier with potential $V_0$.

Note that $w_{-2}/Y_{-2} = \pi/k_{-2}$ and $w_{2}/Y_{2} = \pi/k_{2}$ are the widths of the wells one would deduce by looking at the sinusoidal form of the wavefunctions inside the wells and assuming that they go to zero at the edges.  The fact that the wavefunctions do not vanish at the actual well boundaries for finite $V_{-4}$, $V_0$, and $V_4$ but then approach zero more slowly in those potentials than the sinusoidal wavefunctions do is what is responsible for the factors of $1/U_{-2}$ and $1/U_2$, each of which is larger than unity (but close to unity for small $\alpha_{-3}$, $\alpha{-1}$, $\alpha_1$, and $\alpha_3$).

Combining Eq.\ (\ref{coefrat}) and Eqs.\ (\ref{P}) then gives
\be
\frac{P_L}{P_R} \approx \frac{-(a_{1}-a_{-1})+\sqrt{(a_{1}-a_{-1})^2 + 4p}}
{(a_{1}-a_{-1})+\sqrt{(a_{1}-a_{-1})^2 + 4p}}
= \frac{\sqrt{1+z^2}-z}{\sqrt{1+z^2}+z} = e^{-2r},
\n{Prat}
\ee
where
\be
z \equiv \frac{a_{1}-a_{-1}}{2\sqrt{p}} \equiv \sinh{r}.
\n{z}
\ee


\section{First Excited State Wavefunction and Energy}

If we now turn from the ground state with energy $E_0$ to the first excited state with energy $E_1$, the first excited state wavefunction $\tilde{\psi}(x)$ has the following form in each of the five regions:
\ba
\tilde{\psi}(x) &=& \tilde{\psi}_{-4}(x) = -\tilde{A}_{-4}e^{\tilde{\kappa}_{-4}(x-x_{-3})},\ \ x<x_{-3}, \nonumber\\
\tilde{\psi}(x) &=& \tilde{\psi}_{-2}(x) = -\tilde{A}_{-2}\cos{[\tilde{k}_{-2}(x-\tilde{x}_{-2})]},\ \ x_{-3}<x<x_{-1}, \nonumber\\
\tilde{\psi}(x) &=& \tilde{\psi}_{0}(x)\ \ = \tilde{A}_0\sinh{[\tilde{\kappa}_0(x-\tilde{x}_0)]},\ \ x_{-1}<x<x_1, \nonumber\\
\tilde{\psi}(x) &=& \tilde{\psi}_{2}(x)\ \ = \tilde{A}_2\cos{[\tilde{k}_{2}(x-\tilde{x}_2)]},\ \ x_1<x<x_3, \nonumber\\
\tilde{\psi}(x) &=& \tilde{\psi}_{4}(x)\ \ = \tilde{A}_4e^{\tilde{\kappa}_{4}(x_3-x)},\ \ x_3<x,
\n{psi'}
\ea
where
\ba
\tilde{\kappa}_{-4} &=& [2m(V_{-4}-E_1)]^{1/2}/\hbar, \nonumber\\
\tilde{k}_{-2} &=& [2m(E_1-V_{-2})]^{1/2}/\hbar, \nonumber\\
\tilde{\kappa}_{0} &=& [2m(V_{0}-E_1)]^{1/2}/\hbar, \nonumber\\
\tilde{k}_{2} &=& [2m(E_1-V_{2})]^{1/2}/\hbar, \nonumber\\
\tilde{\kappa}_{4} &=& [2m(V_{4}-E_1)]^{1/2}/\hbar.
\n{mom'}
\ea

Analogous to Eqs.\ (\ref{phi}), it is convenient to define
\ba
\phi_{-3} &\equiv& \pi/2 - \tilde{k}_{-2}(x_{-2}-x_{-3}), \nonumber\\
\phi_{-1} &\equiv& \pi/2 - \tilde{k}_{-2}(x_{-1}-x_{-2}), \nonumber\\
\phi_{1} &\equiv& \pi/2 - \tilde{k}_{2}(x_{2}-x_{1}), \nonumber\\
\phi_{3} &\equiv& \pi/2 - \tilde{k}_{2}(x_{3}-x_{2}).
\n{phi'}
\ea
and to define, instead of Eqs.\ (\ref{y}),
\ba
\tilde{y}_{-2} &\equiv& \left(\frac{E_1-V_{-2}}{K_{-2}}\right)^{1/2}, \nonumber\\
\tilde{y}_{2} &\equiv& \left(\frac{E_1-V_{2}}{K_{2}}\right)^{1/2}.
\n{y'}
\ea
whose determination is equivalent to the determination of the first excited state energy
\be 
E_1 = V_{-2} + K_{-2}\tilde{y}_{-2}^2 = V_{2} + K_{2}\tilde{y}_{2}^2.
\n{E'} 
\ee
Then, instead of Eqs.\ (\ref{s}), define
\ba
\tilde{s}_{-3} &\equiv& \alpha_{-3} \tilde{y}_{-2}, \nonumber\\
\tilde{s}_{-1} &\equiv& \alpha_{-1} \tilde{y}_{-2}, \nonumber\\
\tilde{s}_{1} &\equiv& \alpha_{1} \tilde{y}_{2}, \nonumber\\
\tilde{s}_{3} &\equiv& \alpha_{3} \tilde{y}_{2}.
\n{s'}
\ea
and, instead of Eqs.\ (\ref{r}), define
\ba
\tilde{r}_{-1} &\equiv& \tilde{\kappa}_0(x_0-x_{-1}), \nonumber\\
\tilde{r}_{1} &\equiv& \tilde{\kappa}_0(x_1-x_{0}), \nonumber\\
\tilde{r}_0 &\equiv& \tilde{r}_{-1} + \tilde{r}_1 \equiv \tilde{\kappa}_0(x_1-x_{-1}) \equiv \tilde{\kappa}_0 w_0.
\n{r'}
\ea

The analogue of the matching conditions Eqs.\ (\ref{coef}) are
\ba
\tilde{A}_{-4} &=& \tilde{A}_{-2}\sin{\phi_{-3}}, \nonumber\\
\tilde{A}_{-4}\tilde{\kappa}_{-4} &=& \tilde{A}_{-2}\tilde{k}_{-2}\cos{\phi_{-3}}, \nonumber\\
\tilde{A}_0\sinh{r_{-1}} &=& \tilde{A}_{-2}\sin{\phi_{-1}}, \nonumber\\ 
\tilde{A}_0\tilde{\kappa}_{0}\cosh{r_{-1}} &=& \tilde{A}_{-2}\tilde{k}_{-2}\cos{\phi_{-1}}, \nonumber\\
\tilde{A}_0\sinh{r_1} &=& \tilde{A}_{2}\sin{\phi_1}, \nonumber\\ 
\tilde{A}_0\tilde{\kappa}_{0}\cosh{r_1} &=& \tilde{A}_0\tilde{k}_2\cos{\phi_1}, \nonumber\\
\tilde{A}_{4} &=& \tilde{A}_{2}\sin{\phi_3}, \nonumber\\
\tilde{A}_{4}\tilde{\kappa}_{4} &=& \tilde{A}_{2}\tilde{k}_{2}\cos{\phi_3}.
\n{coef'}
\ea
After eliminating the coefficients $\tilde{A}_{-4}$, $\tilde{A}_{-2}$, $\tilde{A}_0$, $\tilde{A}_{2}$, and $\tilde{A}_{4}$, the analogue of Eqs.\ (\ref{tan}) are
\ba
\tan{\phi_{-3}} &=& \tilde{k}_{-2}/\tilde{\kappa}_{-4}, \nonumber\\
\tan{\phi_{-1}} &=& (\tilde{k}_{-2}/\tilde{\kappa}_{0})\tanh{\tilde{r}_{-1}}, \nonumber\\
\tan{\phi_{1}} &=& (\tilde{k}_{2}/\tilde{\kappa}_{0})\tanh{\tilde{r}_{1}}, \nonumber\\
\tan{\phi_{3}} &=& \tilde{k}_{2}/\tilde{\kappa}_{4}.
\n{tan'}
\ea
These equations then give the following equations that are analogous to Eqs.\ (\ref{set}):
\ba
\sin^{-1}{(\tilde{s}_{-3})} + \sin^{-1}{\left(\frac{\tilde{s}_{-1}}{\sqrt{1+(1-\tilde{s}_{-1}^2)/\sinh^2{(\tilde{r}_{-1})}}}\right)} = \pi - \pi \tilde{y}_{-2}, \nonumber\\
\sin^{-1}{(\tilde{s}_{3})} + \sin^{-1}{\left(\frac{\tilde{s}_{1}}{\sqrt{1+(1-\tilde{s}_{1}^2)/\sinh^2{(\tilde{r}_{1})}}}\right)} = \pi - \pi \tilde{y}_{2}, \nonumber\\
\tilde{r}_{-1} + \tilde{r}_1 = \frac{\pi}{\beta_{-1}}\sqrt{1-\tilde{s}_{-1}^2} = \frac{\pi}{\beta_{1}}\sqrt{1-\tilde{s}_{1}^2}\,.
\n{set'}
\ea
Note the opposite sign in front of the second term inside the square root in the denominators in the first two equations, in comparison with Eqs.\ (\ref{set}).  Then if one writes, analogous to Eqs.\ (\ref{yY}) but with an opposite sign as above,
\ba
\tilde{y}_{-2} &\equiv& Y_{-2}(1+\tilde{\epsilon}_{-2}), \nonumber\\
\tilde{y}_{2} &\equiv& Y_{2}(1+\tilde{\epsilon}_{2}),
\n{yY'}
\ea
the first two of Eqs.\ (\ref{set'}) imply that to first order in $e^{-2\tilde{r}_{-1}}$ and $e^{-2\tilde{r}_{1}}$,
\ba
\tilde{\epsilon}_{-2} &\approx& c_{-2}e^{-2\tilde{r}_{-1}}, \nonumber\\
\tilde{\epsilon}_{2} &\approx& c_{2}e^{-2\tilde{r}_{1}}.
\n{eps'}
\ea
The second two of Eqs.\ (\ref{set'}) imply that to first order in $e^{-2\tilde{r}_{-1}}$ and $e^{-2\tilde{r}_{1}}$,
\ba
\tilde{r}_0 \equiv \tilde{r}_{-1} + \tilde{r}_1 &\approx& a_{-1}-b_{-1}\tilde{\epsilon}_{-2} \approx a_{1}-b_{1}\tilde{\epsilon}_{2} \nonumber\\
&\approx& a_{-1}-b_{-1}c_{-2}e^{-2\tilde{r}_{-1}} \approx a_{1}-b_{1}c_{2}e^{-2\tilde{r}_{1}}.
\n{reps'}
\ea
Then
\ba
\tilde{p} \equiv P e^{-2\tilde{r}_{0}} \approx (a_{-1}-\tilde{r}_0)(a_1-\tilde{r}_0),
\n{p'}
\ea
so
\ba
\tilde{r}_0 \approx \frac{a_{-1}+a_1}{2} - \sqrt{\left(\frac{a_{-1}-a_1}{2}\right)^2 + \tilde{p}}\,.
\n{r_0'}
\ea
As one can do with Eq.\ (\ref{r_0}), one can iterate Eq.\ (\ref{r_0'}) to get a good approximation for $\tilde{r}_0$ and $\tilde{p}$, say starting with the first approximation that $\tilde{p} = 0$, which then gives that the first approximation for $\tilde{r}_0$ is the minimum of $a_{-1}$ and $a_1$ (instead of the maximum, as was the first approximation for $r_0$).  One can then use Eq.\ (\ref{reps'}) to get
\ba
\tilde{\epsilon}_{-2} &\approx& \frac{a_{-1}-\tilde{r}_0}{b_{-1}}
\approx \frac{1}{2b_{-1}}\left[-(a_{1}-a_{-1})+\sqrt{(a_{1}-a_{-1})^2 + 4\tilde{p}}\right], \nonumber\\ 
\tilde{\epsilon}_{2} &\approx& \frac{a_{1}-\tilde{r}_0}{b_{1}}
\approx \frac{1}{2b_{1}}\left[(a_{1}-a_{-1})+\sqrt{(a_{1}-a_{-1})^2 + 4\tilde{p}}\right].
\ea
Then one can plug these back into Eqs.\ (\ref{yY'}) to get $\tilde{y}_{-2}$ and $\tilde{y}_2$ and then insert these into Eqs.\ (\ref{E'}) to get two estimates for the first excited state energy $E_1$, which should be very close to each other and to the exact first excited state energy if $\tilde{r}_{-1}\gg 1$ and $\tilde{r}_1\gg 1$.

As occurs for the ground state wavefunction $\psi(x)$, when the potential is not precisely symmetric about the middle of the barrier, there is an asymmetry in the coefficients $\tilde{A}_{-2}$ and $\tilde{A}_2$ of the first excited state wavefunction $\tilde{\psi}$ in the two wells.  If one uses Eqs.\ (\ref{phi'}), (\ref{r'}), and (\ref{coef'}), and the zeroth-order (in $e^{-2\tilde{r}_{-1}}$ and $e^{-2\tilde{r}_{1}}$) approximations
\ba
\sin{\phi_{-3}} &\approx& \tilde{s}_{-3} \equiv \alpha_{-3} \tilde{y}_{-2} \equiv \alpha_{-3} Y_{-2}(1-\tilde{\epsilon}_{-2}) \approx \alpha_{-3} Y_{-2} \equiv S_{-3}, \nonumber\\
\sin{\phi_{-1}} &\approx& \tilde{s}_{-1} \equiv \alpha_{-1} \tilde{y}_{-2} \equiv \alpha_{-1} Y_{-2}(1-\tilde{\epsilon}_{-2}) \approx \alpha_{-1} Y_{-2} \equiv S_{-1}, \nonumber\\
\sin{\phi_{1}} &\approx& \tilde{s}_{1} \equiv \alpha_{1} \tilde{y}_{2} \equiv \alpha_{1} Y_{2}(1-\tilde{\epsilon}_{2}) \approx \alpha_{1} Y_{2} \equiv S_{1}, \nonumber\\
\sin{\phi_{3}} &\approx& \tilde{s}_{3} \equiv \alpha_{3} \tilde{y}_{2} \equiv \alpha_{3} Y_{2}(1-\tilde{\epsilon}_{2}) \approx \alpha_{3} Y_{2} \equiv S_{3},
\ea
we get
\be
\frac{\tilde{A}_{-2}}{\tilde{A}_2} = \frac{\sinh{\tilde{r}_{-1}\sin{\phi_{-1}}}}{\sinh{\tilde{r}_{1}\sin{\phi_{1}}}} \approx \frac{S_1}{S_{-1}}e^{\tilde{r}_{-1}-\tilde{r}_1}.
\ee
Eqs.\ (\ref{eps'}), (\ref{p'}), and (\ref{r_0'}) then imply that
\be
e^{2(\tilde{r}_{-1}-\tilde{r}_1)} \approx \frac{c_{-2}\tilde{\epsilon}_2}{c_2\tilde{\epsilon}_{-2}} \approx
\left[\frac{b_{-1}c_{-2}}{b_1c_2}\right]
\left[\frac{(a_{1}-a_{-1})+\sqrt{(a_{1}-a_{-1})^2 + 4\tilde{p}}}
{-(a_{1}-a_{-1})+\sqrt{(a_{1}-a_{-1})^2 + 4\tilde{p}}}
\right],
\ee
so
\be
\left(\frac{\tilde{A}_{-2}}{\tilde{A}_2}\right)^2 \approx 
\left[\frac{S_1^2 b_{-1}c_{-2}}{S_{-1}^2 b_1c_2}\right]
\left[\frac{(a_{1}-a_{-1})+\sqrt{(a_{1}-a_{-1})^2 + 4\tilde{p}}}
{-(a_{1}-a_{-1})+\sqrt{(a_{1}-a_{-1})^2 + 4\tilde{p}}}
\right].
\n{coefrat'}
\ee

When one combines Eqs.\ (\ref{P}) with  Eq.\ (\ref{coefrat'}), one gets that for the first excited state wavefunction,
\be
\frac{\tilde{P}_L}{\tilde{P}_R} \approx \frac{(a_{1}-a_{-1})+\sqrt{(a_{1}-a_{-1})^2 + 4\tilde{p}}}
{-(a_{1}-a_{-1})+\sqrt{(a_{1}-a_{-1})^2 + 4\tilde{p}}}
= \frac{\sqrt{1+\tilde{z}^2}+\tilde{z}}{\sqrt{1+\tilde{z}^2}-\tilde{z}} 
= e^{2\tilde{r}} \approx \frac{P_R}{P_L},
\n{Prat'}
\ee
where
\be
\tilde{z} \equiv \frac{a_{1}-a_{-1}}{2\sqrt{\tilde{p}}} \equiv \sinh{\tilde{r}}.
\n{z'}
\ee

The relation given by Eq.\ (\ref{Prat'}), along with the sign change for the excited state wavefunction in the left well in Eqs.\ (\ref{psi'}), is a consequence of the fact that the two energy eigenstates are orthogonal.

\section{Asymmetric Perturbation of the Potential}

Suppose we start off with potential values $V_{-4}$, $V_{-2}$, $V_0$, $V_2$, and $V_4$, and with widths $w_{-2}$, $w_0$, and $w_2$ for the ones that do not extend to infinity as $V_{-4}$ and $V_{4}$ do, such that $a_{-1} = a_1 = a$ and hence so that $P_L/P_R \approx \tilde{P}_R/\tilde{P}_L \approx 1$.  In particular, this implies that in the limit of an infinitely thick barrier between the two wells, the ground state energy of the two wells is the same, say
\be
V_{-2} + K_{-2} Y_{-2}^2 = V_{2} + K_{2} Y_{2}^2 = \bar{E}.
\ee
Then in terms of this common infinitely thick barrier ground state energy $\bar{E}$ of each of the two wells, the dimensionless parameter $a = a_{-1} = a_{1}$ has the value
\be
a = \frac{\pi C_{-1}}{\beta_{-1}} = \frac{\pi C_{1}}{\beta_{1}} = \pi\sqrt{\frac{V_0-\bar{E}}{K_0}}.
\n{a}
\ee

Furthermore, in this case $r_0$ and $p$ are the solutions of the equations
\ba
r_0 &=& a + \sqrt{p} = a + \sqrt{P}\,e^{-r_0},  \nonumber\\
p &=& P e^{-2r_0} = P e^{-2a-2\sqrt{p}}.
\n{pr}
\ea
With our main assumption, which implies that $a \gg 1$, a good approximation is $r_0 \approx a$ and $p \approx P e^{-2a}$.  Perhaps a slightly better approximation is $r_0 \approx a + \sqrt{P}\, e^{-a}$ and $p \approx P e^{-2a -2\sqrt{P}\, e^{-a}}$, though since I am only making approximations to first order in $e^{-a}$, second-order terms in $e^{-a}$ may be comparable to the difference between these two approximations.

Before the perturbation in the potential, but with the finitely thick barrier, to first order in $e^{-a}$ one gets
\ba
E_0 &\approx& \bar{E} - \Delta E, \nonumber\\
E_1 &\approx& \bar{E} + \Delta E,
\n{E0E1nopert}
\ea
with
\be
\Delta E \equiv \frac{1}{2}(E_1-E_0)_\mathrm{unperturbed} \approx \frac{2aK_0}{\pi^2}\sqrt{p}.
\n{DE}
\ee

Now let us perturb the values of the potential slightly and see what effect this has on the ground state and first excited state energies $E_0$ and $E_1$ respectively, as well as on the asymmetry $P_L/P_R \approx \tilde{P}_R/\tilde{P}_L$ of these two wavefunctions.  In principle, we could arbitrarily vary any combination of $V_{-4}$, $V_{-2}$, $V_0$, $V_2$, $V_4$, $w_{-2}$, $w_0$, and $w_2$, but here I shall only make the changes
\ba
V_{-2} &\rightarrow& V_{-2} + \delta V, \nonumber\\
V_{2} &\rightarrow& V_{2} - \delta V,
\n{deltaV}
\ea
for some tiny constant $\delta V$.  It is also convenient to define
\be
v \equiv \frac{\delta V}{\Delta E}
\n{v}
\ee
as a dimensionless measure of the perturbation of the potential.  It is only necessary that $\delta V \equiv \Delta E\,v$ be much smaller than the potential differences $W_{-3}$, $W_{-1}$, $W_1$, and $W_3$, but since $\Delta E$ typically is much smaller than all of these, it is not necessary that $v$ itself be small compared with unity; the analysis below is valid even for $v \gg 1$.

This highly restricted perturbation will not change quantities such as $K_{-2}$, $K_0$, and $K_2$, but it will change quantities such as $\alpha_{-3}$, $\alpha_{-1}$, $\alpha_{1}$, $\alpha_{3}$, $\beta_{-1}$, $\beta_{1}$, $Y_{-2}$, $Y_{2}$, $S_{-2}$, and $S_{2}$ to $\alpha_{-3} + \delta\alpha_{-3}$ etc., where after the change, any of the quantities that are written without a $\delta$ in front of it, are the values corresponding to the unperturbed potential that gives $a_{-1} = a_1 = a$, and all of the quantities with $\delta$ written in front of it, such as $\delta\alpha_{-3}$, are perturbations that will be calculated to first order in $\delta V$.  For example, one gets that, to first order in the potential perturbation $\delta V \equiv \Delta E\, v$,
\ba
\frac{\delta\alpha_{-3}}{\alpha_{-3}} &=& +\frac{\delta V}{2W_{-3}}, \nonumber\\
\frac{\delta\alpha_{-1}}{\alpha_{-1}} &=& \frac{\delta\beta_{-1}}{\beta_{-1}}
= +\frac{\delta V}{2W_{-1}}, \nonumber\\
\frac{\delta\alpha_{1}}{\alpha_{1}} &=& \frac{\delta\beta_{1}}{\beta_{1}}
= -\frac{\delta V}{2W_{1}}, \nonumber\\
\frac{\delta\alpha_{3}}{\alpha_{3}} &=& -\frac{\delta V}{2W_{3}}, \nonumber\\
\frac{\delta Y_{-2}}{Y_{-2}} &=& 
-\frac{1}{2\pi}U_{-2}\left(\frac{T_{-1}}{W_{-1}}+\frac{T_{-3}}{W_{-3}}\right) \delta V, \nonumber\\
\frac{\delta Y_{2}}{Y_{2}} &=& 
+\frac{1}{2\pi}U_{2}\left(\frac{T_{1}}{W_{1}}+\frac{T_{3}}{W_{3}}\right)
\delta V, \nonumber\\
\frac{\delta S_{-3}}{S_{-3}} &=& 
-\frac{1}{2\pi}U_{-2}\left(\frac{T_{-1}}{W_{-1}}-\frac{T_{-1}+\pi Y_{-2}} {W_{-3}}\right) \delta V, \nonumber\\
\frac{\delta S_{-1}}{S_{-1}} &=& 
-\frac{1}{2\pi}U_{-2}\left(\frac{T_{-3}}{W_{-3}}-\frac{T_{-3}+\pi Y_{-2}} {W_{-1}}\right) \delta V, \nonumber\\
\frac{\delta S_{1}}{S_{1}} &=& 
+\frac{1}{2\pi}U_{2}\left(\frac{T_{3}}{W_{3}}-\frac{T_{3}+\pi Y_{2}} {W_{1}}\right) \delta V, \nonumber\\
\frac{\delta S_{3}}{S_{3}} &=& 
+\frac{1}{2\pi}U_{2}\left(\frac{T_{1}}{W_{1}}-\frac{T_{1}+\pi Y_{2}} {W_{3}}\right) \delta V, \nonumber\\
\frac{\delta a_{-1}}{a} &=&
-\frac{1}{2\pi}U_{-2}\left(\frac{S_{-1}C_{-1}+T_{-3}+\pi Y_{-2}}
{C_{-1}^2 W_{-1}}-\frac{T_{-1}^2T_{-3}}{W_{-3}}\right) \delta V, \nonumber\\
\frac{\delta a_{1}}{a} &=&
+\frac{1}{2\pi}U_{2}\left(\frac{S_{1}C_{1}+T_{3}+\pi Y_{2}}
{C_{1}^2 W_{1}}-\frac{T_{1}^2T_{3}}{W_{3}}\right) \delta V.
\ea

Now if one defines
\be
F \equiv \frac{1}{2\pi}\left[U_2\left(S_1^2T_1+S_3^2T_3\right)
-U_{-2}\left(S_{-1}^2T_{-1}+S_{-3}^2T_{-3}\right)\right],
\n{F}
\ee
\be
G \equiv 1 - \frac{1}{2\pi}\left[U_2\left(S_1^2T_1+S_3^2T_3\right)
+U_{-2}\left(S_{-1}^2T_{-1}+S_{-3}^2T_{-3}\right)\right],
\n{G}
\ee
then in the infinitely thick barrier limit the energies of the particles in the left and right wells after the perturbation of the potential are, to first order in the perturbation $\delta V \equiv \Delta E\, v$,
\ba
E_L &=& V_{-2}+\delta V + K_{-2}(Y_{-2}+\delta Y_{-2})^2 
\approx \bar{E} + \Delta E\,(Fv + Gv), \nonumber\\
E_R &=& V_{2}-\delta V + K_{2}(Y_{2}+\delta Y_{2})^2 
\approx \bar{E} + \Delta E\,(Fv - Gv).
\ea

Furthermore, since before the perturbation $a_1-a_{-1} = 0$, after the perturbation one has
\be
a_1-a_{-1} \approx \frac{\pi^2 G}{a K_0} \delta V \approx 2G\sqrt{p}\,v.
\ee
Then from the definitions Eqs.\ (\ref{z}) and (\ref{z'}) with $p = \tilde{p}$ taken to be the value from the unperturbed potential,
\ba
z &\equiv& \sinh{r} \equiv \frac{a_1-a_{-1}}{2\sqrt{p}} \approx \tilde{z}
\approx \frac{\pi^2 G}{2aK_0\sqrt{p}}\,\delta V \approx Gv, \nonumber\\
r &\equiv& \frac{1}{2}\ln{\left(\frac{\sqrt{1+z^2}+z}{\sqrt{1+z^2}-z}\right)}
= \ln{(\sqrt{1+z^2}+z)} \approx \tilde{r}
\approx  \frac{1}{2}\ln{\frac{P_R}{P_L}}
\approx  \frac{1}{2}\ln{\frac{\tilde{P}_L}{\tilde{P}_R}}.
\ea
Then the ground state energy $E_0$ and first excited state energy $E_1$ will have the values, to first order in $\delta V \equiv \Delta E\, v$,
\ba
E_0 &\approx& \bar{E} + \Delta E\,(Fv - \sqrt{1+G^2v^2}), \nonumber\\
&\approx& \bar{E} + \Delta E\,(Fv - \sqrt{1+z^2}), \nonumber\\
&=& \bar{E} + \Delta E\,(Fv - \cosh{r}), \nonumber\\
&\approx& \bar{E}+\Delta E\,\left(Fv
- \frac{1}{2} \sqrt{\frac{P_R}{P_L}} 
- \frac{1}{2} \sqrt{\frac{P_L}{P_R}}\right),
\nonumber\\
E_1 &\approx& \bar{E} + \Delta E\,(Fv + \sqrt{1+G^2v^2}), \nonumber\\
&\approx& \bar{E} + \Delta E\,(Fv + \sqrt{1+z^2}), \nonumber\\
&=& \bar{E} + \Delta E\,(Fv + \cosh{r}), \nonumber\\
&\approx& \bar{E}+\Delta E\,\left(Fv
+ \frac{1}{2} \sqrt{\frac{P_R}{P_L}} 
+ \frac{1}{2} \sqrt{\frac{P_L}{P_R}}\right),
\ea

In terms of the ratio $P_L/P_R \approx e^{-2r} = (\sqrt{1+z^2}-z)/(\sqrt{1+z^2}+z)$ of the probabilities that a particle in the ground state wavefunction is on the left side and on the right side of the location of the minimum of the wavefunction inside the barrier between the two wells, one can also express the perturbation in the potential as
\be
\delta V \approx \frac{\Delta E}{2G}
\left(\sqrt{\frac{P_R}{P_L}} - \sqrt{\frac{P_L}{P_R}}\right)
\approx \frac{E_1-E_0}{2G}\left(\frac{P_R-P_L}{P_R+P_L}\right).
\n{dV}
\ee
Alternatively, when $\delta V > 0$ so that $P_R > P_L$, the energy difference between the ground state and the first excited state is
\be
E_1 - E_0 \approx 2G\left(\frac{P_R+P_L}{P_R-P_L}\right)\delta V.
\ee

In the limit that $P_R \gg P_L$, one gets $E_1 - E_0 \approx 2G\delta V$.  If one further takes the limit that $\bar{E} - V_{-2}$ is much smaller than both $W_{-3}$ and $W_{-1}$ and that $\bar{E} - V_{2}$ is much smaller than both $W_{1}$ and $W_{3}$, then $G \approx 1$, so that then when $P_R \gg P_L$, one gets $E_1 - E_0 \approx 2\delta V$.  In the symmetric unperturbed case in which $V_{-4} = V_4$, $V_{-2} = V_2$, and $w_{-2} = w_2$ before the perturbation, with these limits one gets that the excitation energy of the first excited state, $E_1 - E_0$, is approximately simply the potential difference between the perturbed well potentials $V_{-2}+\delta V$ and $V_2-\delta V$.  In this set of limits, raising $V_{-2}$ by $\delta V$ produces the same rise in the energy $E_1$ of the excited state wavefunction (which is then almost entirely in the left well), and lowering $V_{2}$ by $\delta V$ produces the same lowering of the energy $E_0$ of the ground state wavefunction (which is then almost entirely in the right well).  However, outside these limits, the two wavefunctions are not almost entirely within the two wells, and the situation is more complicated.

One can get the ground state wavefunction $\psi(x)$ of energy $E_0$ and the first excited state wavefunction $\tilde{\psi}(x)$ of energy $E_1$ for the finitely thick barrier as approximate superpositions of the normalized ground state wavefunctions $\psi_L(x)$ and $\psi_R(x)$ of the two wells in the infinitely thick barrier limit:
\ba
\psi(x) &\approx& \sqrt{P_L}\, \psi_L(x) + \sqrt{P_R}\, \psi_R(x), \nonumber\\
\tilde{\psi}(x) &\approx& -\sqrt{\tilde{P}_L}\, \psi_L(x)
 + \sqrt{\tilde{P}_R}\, \psi_R(x) \nonumber\\
&\approx& -\sqrt{{P}_R}\, \psi_L(x) + \sqrt{{P}_L}\, \psi_R(x),
\n{wavefunctions}
\ea
where
\ba
P_L = \tilde{P}_R \approx \frac{e^{-r}}{e^r+e^{-r}}
=\frac{\sqrt{1+z^2}-z}{2\sqrt{1+z^2}}, \nonumber\\
P_R = \tilde{P}_L \approx \frac{e^{r}}{e^r+e^{-r}}
=\frac{\sqrt{1+z^2}+z}{2\sqrt{1+z^2}}.
\n{PLPR}
\ea

In terms of the two basis states $\psi_L(x)$ and $\psi_R(x)$ of the two wells in the infinitely thick barrier limit, the Hamiltonian $H$ is given approximately by the following $2\times 2$ matrix:
\be
H \approx \left( \begin{array}{cc}
E_L & -\Delta E \\
-\Delta E & E_R
\end{array} \right).
\ee
Then one gets the two eigenstates of this Hamiltonian as
\be
\begin{array}{cccc}
\left( \begin{array}{cc}
E_L & -\Delta E \\
-\Delta E & E_R
\end{array} \right)\!\!\!\!\!
&
\left(\!\! \begin{array}{c}
\sqrt{P_L} \\
\sqrt{P_R}
\end{array} \!\!\right)
&
\approx \, E_0\!\!\!\!
&
\left(\!\! \begin{array}{c}
\sqrt{P_L} \\
\sqrt{P_R}
\end{array} \!\!\right),
\end{array}
\ee 
\be
\begin{array}{cccc}
\left( \begin{array}{cc}
E_L & -\Delta E \\
-\Delta E & E_R
\end{array} \right)\!\!\!\!\!
&
\left(\!\! \begin{array}{c}
-\sqrt{P_R} \\
\sqrt{P_L}
\end{array} \!\!\right)
&
\approx \, E_1\!\!\!\!
&
\left(\!\! \begin{array}{c}
-\sqrt{P_R} \\
\sqrt{P_L}
\end{array} \!\!\right).
\end{array}
\ee 

\section{Explicit Results for a Simple Example}

Let us consider an example in which for simplicity we set $\hbar = 1$ and $m = 2$.  Let the unperturbed potential values be $V_{-4} = V_0 = V_4 = 1$ and $V_{-2} = V_2 = 0$, so that each of the two wells has zero potential and the external region and the barrier between the two wells each has unit potential.  Then $W_{-3} \equiv V_{-4}-V_{-2} = W_{-1} \equiv V_0-V_{-2} = W_1 \equiv V_0-V_2 = W_3 \equiv V_4-V_2 = 1$.  Let the widths of the wells be $w_{-2} = w_2 = 2\pi/3$ and the width of the barrier between the two wells be $w_0 = 10\pi/3 = 5 w_{-2} = 5 w_2$.  Then $K_{-2} = K_2 = 9/16 = 25K_0$ and $K_0 = 9/400 = 0.04 K_{-2} = 0.04 K_2$, so the quantities that are dimensionless parameters (even without choosing the units so that $\hbar = 1$, $m=2$, and so that the potential differences are all unity) are $\alpha_{-3} = \alpha_{-1} = \alpha_1 = \alpha_3 = 3/4 \equiv \alpha$ and $\beta_{-1} = \beta_1 = \alpha/5 = 3/20 \equiv \beta$.  From these, one can also get $\gamma_{-3} = \gamma_{-1} = \gamma_1 = \gamma_3 = (\pi\alpha)/(\pi+2\alpha) \equiv \gamma$.

Then in the infinitely thick barrier limit (taking $w_0$ to infinity instead of using the finite value given above), Eqs.\ (\ref{Y}) have the simple solution (used to set the parameters above) $Y_{-2} = Y_2 = 2/3 \equiv Y$, then giving $\Phi_{-3} = \Phi_{-1} = \Phi_1 = \Phi_3 = \pi/6 \equiv \Phi$, $S_{-3} = S_{-1} = S_1 = S_3 = 1/2 \equiv S \equiv \alpha Y \equiv \sin{\Phi}$, $C_{-3} = C_{-1} = C_1 = C_3 = \sqrt{3}/2 \equiv \cos{\Phi}$, and $T_{-3} = T_{-1} = T_1 = T_3 = \sqrt{3}/3 \equiv \tan{\Phi}$.  The energy of the ground state in each well in the infinitely thick barrier limit is then $\bar{E} = V_2 + K_2 Y_2^2 = (4/9)K_2 = 1/4 = 0.25$. 

One then gets
\ba
a &=& a_{-1} = a_1 = \frac{\pi C}{\beta} = \frac{10\pi}{\sqrt{3}} = 18.137\,993\,6423, \nonumber\\
b &=& b_{-1} = b_1 = \frac{\pi S T}{\beta} = T^2 a = \frac{10\pi}{3\sqrt{3}} = 6.045\,997\,880\,78, \nonumber\\
U &=& \frac{3\pi}{2\pi+2\sqrt{3}} = 0.966\,912\,950\,84, \nonumber\\
c &=& \frac{2}{\pi}SCU = 0.266\,543\,524\,679, \nonumber\\
P &=& b^2c^2 = \left(\frac{5\pi}{2\pi+2\sqrt{3}}\right)^2 = 2.597\,001\,818\,08, \nonumber\\
\gamma &=& \frac{3\pi}{4\pi+6} = 0.507\,626\,296\,843. \nonumber\\
F &=& 0,\nonumber\\
G &=& \frac{4\pi+3\sqrt{3}}{4\pi+4\sqrt{3}} = 0.911\,152\,158\,473.
\n{abc}
\ea

When one combines Eqs.\ (\ref{Yseries}) and (\ref{S}) with these parameter values, one gets the following truncated series approximation for the exact solution $S=0.5$:
\ba
S\!\! &\approx&\!\! \gamma\! -\! \frac{1}{3\pi}\gamma^4\! -\! \frac{3}{20\pi}\gamma^6 \!+\! \frac{1}{3\pi^2}\gamma^7 \!-\! \frac{5}{56\pi}\gamma^8 \!+\! \frac{2}{5\pi^2}\gamma^9
\!-\!\left(\frac{35\pi^2+256}{{576\pi^3}}\right)\gamma^{10} 
\!+\! \frac{689}{1680\pi^2}\gamma^{11} \nonumber\\
&=& \!\! 0.500\,008\,388\,946,
\n{Sseries}
\ea
which has a relative error of $1.677\,789\times 10^{-5}$, which does not appear too bad for $\alpha = 0.75$, which is not that small.  On the other hand, the first estimate given by Eqs.\ (\ref{Yest1}) is $Y_{-2(1)} = Y_{2(1)} = Y_{(1)} = 0.667\,441\,203\,024$ and leads to a corresponding first estimate for $S = \alpha Y$ of $S_{(1)} = 0.500\,580\,902\,268$, with relative error $1.116\,180\,454\times 10^{-3}$, which is already rather small.  Then one iteration of Eqs.\ (\ref{Yestn}) leads to $S_{(2)} = 0.500\,000\,040\,032$, with relative error $8.0064\times 10^{-8}$, already smaller than that of the truncated series, and a second iteration leads to $S_{(3)} = 0.5$, the correct answer for $S=1/2$ to all 12 digits given on my pocket calculator.

For solving Eqs.\ (\ref{pr}), a first approximation is that $r_0 \approx a = (10\pi)/\sqrt{3} = 18.137\,993\,6423$ and that $p \approx Pe^{-2a} = [5\pi/(2\pi+2\sqrt{3})]\exp{(-20\pi/\sqrt{3})} = \\ 4.570\,999\,253\,12\times 10^{-16}$.  Then a second approximation is that (using this approximation for $p$) $r_0 = a + \sqrt{p} = 18.137\,993\,6637$ and that $p \approx Pe^{-2r_0} = 4.570\,999\,057\,95\times 10^{-16}$.  Inserting these back into the equations gave a third approximation the same as the second one to all 12 digits given by my pocket calculator.

Then $\sqrt{p} = 2.137\,989\,489\,67\times 10^{-8}$, so that half the energy splitting between the ground state energy $E_0$ and the first excited state energy $E_1$ for the unperturbed potential is
\be
\Delta E \approx \frac{2aK_0}{\pi^2}\sqrt{p} = \frac{3\sqrt{3}}{20\pi}\sqrt{p}
= 1.768\,103\,075\,65\times 10^{-9} = 7.072\,412\,302\,58\times 10^{-9} \bar{E}.
\n{DeltaE}
\ee
Therefore, the ground state and first excited state energies for the unperturbed potential are
\ba
E_0 &=& \bar{E} - \Delta E = 0.999\,999\,992\,928 \bar{E}, \nonumber\\
E_1 &=& \bar{E} + \Delta E = 1.000\,000\,007\,072 \bar{E}.
\ea

Now suppose we perturb the potentials in the two wells (zero initially) so that after the perturbation they become
\ba
V_{-2} &=& \delta V = \Delta E\,v, \nonumber\\
V_{2} &=& -\delta V = -\Delta E\,v.
\ea
Then the ground state and first excited state energies for the perturbed potential become
\ba
E_0 &=& \bar{E} - \Delta E\,\sqrt{1+G^2v^2}, \nonumber\\
E_1 &=& \bar{E} + \Delta E\,\sqrt{1+G^2v^2},
\ea
and the ratio of the probability for the particle in the ground state to be on the right side to that on the left side is
\be
\frac{P_R}{P_L} = \frac{\sqrt{1+G^2v^2}+Gv}{\sqrt{1+G^2v^2}-Gv}.
\ee

For example, if $v = 1$ or $\delta V = \Delta E$, so that the potential $V_{-2}$ in the left well is the same as the excess excited state energy $E_1 - \bar{E}$ for the unperturbed potential, and so that the right well potential $V_{2}$ equals $E_0 - \bar{E}$ for the unperturbed potential, then
\ba
E_0 &=& \bar{E} - \Delta E\,\sqrt{1+G^2} = 0.999\,999\,990\,432 \bar{E}, \nonumber\\
E_1 &=& \bar{E} + \Delta E\,\sqrt{1+G^2} = 1.000\,000\,009\,568 \bar{E},
\ea
and the ratio of probabilities is
\be
\frac{P_R}{P_L} = \frac{\sqrt{1+G^2}+G}{\sqrt{1+G^2}-G} = 5.125\,697\,629\,24.
\ee
Alternatively, if $v = 2$ or $\delta V = 2\Delta E$, then
\ba
E_0 &=& \bar{E} - \Delta E\,\sqrt{1+4G^2} = 0.999\,999\,985\,299 \bar{E}, \nonumber\\
E_1 &=& \bar{E} + \Delta E\,\sqrt{1+4G^2} = 1.000\,000\,014\,701 \bar{E},
\ea
and the ratio of probabilities is
\be
\frac{P_R}{P_L} = \frac{\sqrt{1+4G^2}+2G}{\sqrt{1+4G^2}-2G} = 15.217\,458\,0971.
\ee

Taking a potential perturbation that is rather large in comparison with $\Delta E$ but quite small in comparison with $\bar{E}$, let $\delta V = 0.000001 \bar{E} = 0.000\,000\,25$, the last number being expressed in the units in which the unperturbed potential differences were unity.  Then $v \equiv \delta V/\Delta E = 141.394\,471\,534$, $Gv = 128.831\,877\,934$, $\sqrt{1+G^2v^2} = 128.835\,758\,903$, and $\sqrt{1+G^2v^2}\Delta E = 0.000\,000\,911\,179\,606\,278 \bar{E}$, so
\ba
E_0 &=& \bar{E} - \Delta E\,\sqrt{1+G^2v^2} = 0.999\,999\,088\,820 \bar{E}, \nonumber\\
E_1 &=& \bar{E} + \Delta E\,\sqrt{1+G^2v^2} = 1.000\,000\,911\,180 \bar{E},
\ea
and the ratio of probabilities is
\be
\frac{P_R}{P_L} = \frac{\sqrt{1+G^2v^2}+Gv}{\sqrt{1+G^2v^2}-Gv} = 66\,393.
\ee
Therefore, even though the energy levels are perturbed from their mean by less than one part in million, the ratio of probabilities is enormous, nearly two thirds of a lakh (the number used in India for 100\,000).

This feature seems to be an extreme limit (with only two sites) of one aspect of Anderson localization \cite{A}, the phenomenon that when a potential is not sufficiently homogeneous under discrete translations across the system, the energy eigenstates are each concentrated on only a fraction of the total system.

In conclusion, we have explicit formulas and iteration algorithms (requiring only a very small number of iterations) for getting highly accurate energies and wavefunctions for the ground state and first excited state of a double square well potential of a general form when the barrier between the two wells is sufficiently thick.  We also have explicit formulas for how the energies and wavefunctions change under a simple small change in the potentials.  A change in the potentials that is very small with respect to the potential differences in the different regions can give a huge change in the ratio of probabilities for the particle to be on the left and right side of the barrier between the two wells, expressing by explicit formulas the beautiful numerical results of \cite{DM}. 

\section{Acknowledgments}

This paper was motivated by discussions with Frank Marsiglio (as well as by his papers \cite{JM,DM}), who also offered advice on my calculations, independent derivations of some of them, and suggestions for the manuscript.  This work was supported by the Natural Sciences and Engineering Council of Canada.

\end{document}